# Superconductivity in the metastable 1T' and 1T''' phases of MoS$_2$ crystals

C. Shang[1,†], Y. Q. Fang[2,3,†], Q. Zhang[1], N. Z. Wang[1], Y. F. Wang[1], Z. Liu[1], B. Lei[1], F. B. Meng[1], L. K. Ma[1], T. Wu[1,6,7,8], Z. F. Wang[1], C. G. Zeng[1,5], F. Q. Huang[2,3,6], Z. Sun[1,4,6 *], and X. H. Chen[1,6,7,8 *]

1. Key Laboratory of Strongly-coupled Quantum Matter Physics, Chinese Academy of Sciences, and Hefei National Laboratory for Physical Sciences at Microscale, and Department of Physics, University of Science and Technology of China, Hefei, Anhui 230026, China

2. State Key Laboratory of High Performance Ceramics and Superfine Microstructure, Shanghai Institute of Ceramics, Chinese Academy of Sciences, Shanghai 200050, China

3. State Key Laboratory of Rare Earth Materials Chemistry and Applications, College of Chemistry and Molecular Engineering, Peking University, Peking 100871, China

4. National Synchrotron Radiation Laboratory, University of Science and Technology of China, Hefei, Anhui 230026, China

5. Synergetic Innovation Center of Quantum Information and Quantum Physics, University of Science and Technology of China, Hefei, Anhui 230026, China

6. CAS Center for Excellence in Superconducting Electronics (CENSE), Shanghai 200050, China

7. CAS Center for Excellence in Quantum Information and Quantum Physics, Hefei, Anhui 230026, China

8. Collaborative Innovation Center of Advanced Microstructures, Nanjing University, Nanjing 210093, China

†These authors contributed equally to this work.

Transition-metal dichalcogenides open novel opportunities for the exploration of exciting new physics and devices. As a representative system, 2H-MoS$_2$ has been extensively investigated owing to its unique band structure with a large band gap, degenerate valleys and non-zero Berry curvature. However, experimental studies of metastable 1T polytypes have been a challenge for a long time, and electronic properties are obscure due to the inaccessibility of single phase without the coexistence of 1T', 1T'' and 1T''' lattice structures, which hinder the broad applications of MoS$_2$ in future nanodevices and optoelectronic devices. Using K$_x$(H$_2$O)$_y$MoS$_2$ as the precursor, we have successfully obtained high-quality layered crystals of the metastable 1T'''-MoS$_2$ with $\sqrt{3}a \times \sqrt{3}a$ superstructure and metastable 1T'-MoS$_2$ with $a \times 2a$ superstructure, as evidenced by structural characterizations through scanning tunneling microscopy, Raman spectroscopy and X-ray diffraction. It is found that the metastable 1T'-MoS$_2$ is a superconductor with onset transition temperature (T$_c$) of 4.2 K, while the metastable 1 T'''-MoS$_2$ shows either superconductivity with T$_c$ of 5.3 K or insulating behavior, which strongly depends on the synthesis procedure. Both of the metastable polytypes of MoS$_2$ crystals can be transformed to the stable 2H phase with mild annealing at about 70°C in He atmosphere. These findings provide pivotal information on the atomic configurations and physical properties of 1T polytypes of MoS$_2$.

PACS numbers: 74.25.F-, 74.25.-q, 74.70.Xa

The transition-metal dichalcogenides (TMDs) show diverse physical properties, including metallic, semi-metallic, superconducting, semiconducting, Mott transition or charge density wave subjects, which vary drastically depending on the composition, structure and dimensionality. Owing to their versatile physical properties rooted in the unique crystal structures as well as electronic structures, TMDs have come to the forefront of intense studies of two-dimensional materials and have clearly demonstrated their potential in modern science and technology.[1-5] For instance, the existence of a direct gap in monolayer, together with the large exciton energy and strong spin-valley coupling, provides novel opportunities for intriguing devices.[6-10] New physics and exotic functions may also emerge upon combining different layers,[11,12] each of which has distinct physical properties. Moreover, these layered materials can serve as excellent hosting systems for ion insertion,[13-15] which can further tune their electronic properties.

In the TMD family, there are many metastable materials.[16-19] Contrary to the extensively studied TMDs with stable structures, the preparation, structural characterization and investigation of physical properties of metastable TMDs have been a challenge in the long history, because they usually mix with stable TMD polytypes and are highly susceptible to structural transitions. However, recent studies suggest that metastable TMDs can host novel electronic properties. For instance, owing to the band inversion and a small band gap induced by spin-orbit interactions, some group VI TMDs with metastable 1T' polytype are predicted to have quantum spin Hall effect.[20] Theoretical calculations have also indicated that, associated with particular types of atomic coordinates, 1T-$MoS_2$ could become a metal, a normal insulator, a ferroelectric insulator,[21,22] etc. $MoS_2$ transistors based on 1T/2H interface have demonstrated excellent performance.[23-25]

As a model system, 2H-MoS$_2$ has been intensively investigated to explore the underlying physics and potential applications. In contrast to the highly stable 2H polymorph, 1T-MoS$_2$ is metastable and susceptible to the transition to 2H. In earlier studies, transmission electron microscope (TEM) and scanning tunneling microscopy (STM) have detected small 1T-MoS$_2$ patches that coexist with 2H polytype.[26,27] Nevertheless, physical properties of metastable 1T-MoS$_2$ have not been clearly characterized by various probing techniques. Experimental investigation has been obstructed by the lack of high-quality samples, though theoretical studies suggest there exists exotic physics. Recently, high activity of 1T'-MoS$_2$ crystals for hydrogen evolution reaction has been demonstrated.[28] Moreover, some trace of superconductivity has been revealed in the polycrystalline 1T and 1T' polytypes.[29,30] However, owing to the fact that multiple metastable phases and impurities could exist in the powder samples, the electronic properties and crystal structures of metastable polytypes require further investigation by means of comprehensive transport, magnetic and structural characterizations. In this paper, we show that we have successfully synthesized high-quality layered crystals of metastable 1T' and 1T'''-MoS$_2$ with improved preparation method. The particular superlattices of $a \times 2a$ (1T') and $\sqrt{3}a \times \sqrt{3}a$ (1T''') have been unambiguously identified using XRD, Raman, and STM. Transport measurements show superconductivity with T$_c$ = 4.2 and 5.3 K in metastable 1T' and 1T''' polytypes, respectively. By regulating the oxidation procedure, we have also synthesized 1T'''-MoS$_2$ insulator. The metastable lattice structures can be easily converted into 2H polytype by mild annealing at about 70º C in He atmosphere. Our study demonstrates that systematic investigations of metastable TMDs can reveal more novel physical properties.

In the studies of MoS$_2$, three types of superlattice in metastable 1T-MoS$_2$ have been reported, namely, a × 2a (1T'), $2a \times 2a$ (1T") and $\sqrt{3}a \times \sqrt{3}a$ (1T''') (see Fig. 1). The variations among them are related to synthesis procedure.[28-32] The lattice distortion in MoS$_2$ layers can be induced by extra electrons donated from the intercalated alkali metals, and various superstructures can be induced in 1T-MoS$_2$ after the removal of extra electrons during oxidation process. To obtain layered crystals of metastable 1T' and 1T'''-MoS$_2$, we first reacted KS, MoS$_2$ and Mo with a mass ratio of KS : MoS$_2$ : Mo = 142 : 160 : 96 in a sealed and evacuated quartz tube to prepare the starting material KMoS$_2$. The mixture was heated to 1000 °C in 1000 minutes and kept at this temperature for 2000 minutes, and then slowly cooled down to 750 °C in 1000 minutes, and then the furnace was powered off and the temperature decreased gradually down to room temperature. The resulting KMoS$_2$ was put into a beaker with deionized water, and then we acquired K$_x$(H$_2$O)$_y$MoS$_2$ crystals, as confirmed by the XRD pattern in Fig. 1(c), which has a monoclinic lattice with a = 5.69 Å, b = 3.24 Å, c = 9.40 Å, β = 100.52°.[32] Using K$_x$(H$_2$O)$_y$MoS$_2$ layered crystals as the precursor, we apply K$_2$Cr$_2$O$_7$ to regulate the oxidation procedure, and eventually we are able to acquire layered crystals of 1T polytypes with a size of about 50 μm. The resulting products vary drastically with different mass ratio of K$_x$(H$_2$O)$_y$MoS$_2$ and K$_2$Cr$_2$O$_7$, and we can obtain 1T'-MoS$_2$ superconductor, 1T'''-MoS$_2$ superconductor and 1T'''-MoS$_2$ insulator. The detailed information is presented in the Supplemental Material.

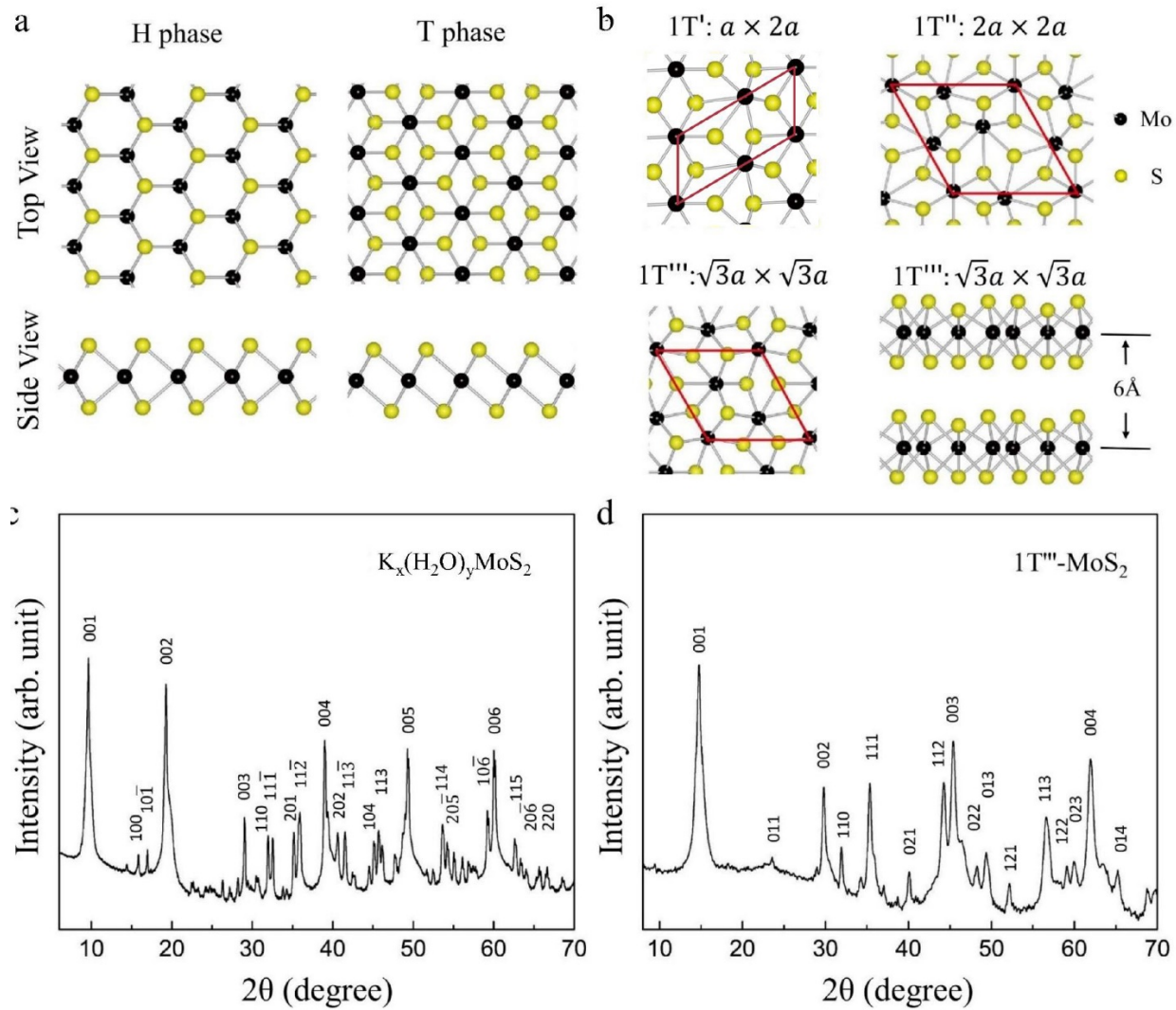

**Figure 1.** The crystal structures of MoS$_2$. (a) The top and side views of H and T phases of MoS$_2$ in one monolayer. (b) Different lattice structures of 1T', 1T'' and 1T'''-MoS$_2$. The red lines indicate the corresponding superstructures of a × 2a, $2a \times 2a$ and $\sqrt{3}a \times \sqrt{3}a$, respectively. (c) X-ray diffraction pattern of K$_x$(H$_2$O)$_y$MoS$_2$, which has a monoclinic lattice with a = 5.69 Å, b = 3.24 Å, c = 9.40 Å, β = 100.52° and is consistent with reports in ref. 32. (d) X-ray diffraction pattern of 1T'''-MoS$_2$, which has a trigonal lattice with a = 5.60 Å, c = 5.98 Å, α = 90°, β = 90° and γ = 120°.

We first focus on 1T'''-MoS$_2$, the physical properties of which have been unclear, because the pure phase of this material has not be synthesized before. Though theoretical calculations have been performed to investigate this metastable material and predict it is an insulator,[22] their electronic properties have yet to be experimentally explored. Fig. 1(d) shows the XRD pattern of 1T'''-MoS$_2$ (a mixture of both the superconducting and insulating compounds). By comparing with the standard 2H and 1T' lattice structures, we have not found evident diffraction peaks from 2H-MoS$_2$, 1T'-MoS$_2$ as well as other impurities, which indicates the high purity of the 1T''' layered crystals. According to the XRD measurement, we can determine the lattice parameters, a = 5.60 Å, c = 5.98 Å, α = 90°, β = 90° and γ = 120°, being consistent with the reported values.[32] More importantly, the as-prepared layered crystals are suitable for various experimental characterizations to reveal the intrinsic physical properties. However, we stress that the difference of crystal structures of 1T'''-MoS$_2$ superconductor and insulator cannot be resolved by XRD, STM, and Raman measurements (see the Supplemental Material for more details), and more high-resolution structural characterizations are required to examine the difference between these two phases in further studies. Here we focus on the superconducting 1T'''-MoS$_2$ crystal, while the transport and structural characterizations on the insulating compound are presented in the Supplemental Material.

Figure 2(a) shows a typical crystal of 1T'''-MoS$_2$ superconductor and the configuration for the four-probe resistance measurements. Atomic-resolution STM measurements on these crystals clearly illustrate the $\sqrt{3}a \times \sqrt{3}a$ superstructure of 1T''' polytype (see Fig. 2(b)). Raman spectra have been intensively utilized in our studies to distinguish different polytypes and examine the quality of each sample. In Fig. 2(c), we show typical Raman spectra taken from 1T'''-MoS$_2$ superconductor. To verify that the Raman modes we obtained are associated with the intrinsic lattice structure of 1T''' polytype, we have performed density functional theory (DFT) calculations to simulate the Raman data. A Gaussian broadening was applied to the DFT spectrum with σ = 3 cm$^{-1}$. As shown in Fig. 2(c), there is a good consistency between DFT calculation and experimental data, indicating that Raman is a reliable tool to characterize the lattice structure of 1T''' polytype. To examine the quality of the samples used for transport measurements, we have carried out Raman measurements with a spatial resolution of 1 μm to collect data over an area of 10 μm × 20 μm covering the four electrodes of 1T'''-MoS$_2$ (Fig. 2(d)). The Raman data taken from different spots are highly consistent with each other, indicating that the quality of the 1T''' lattice structure is uniform at micrometer scale.

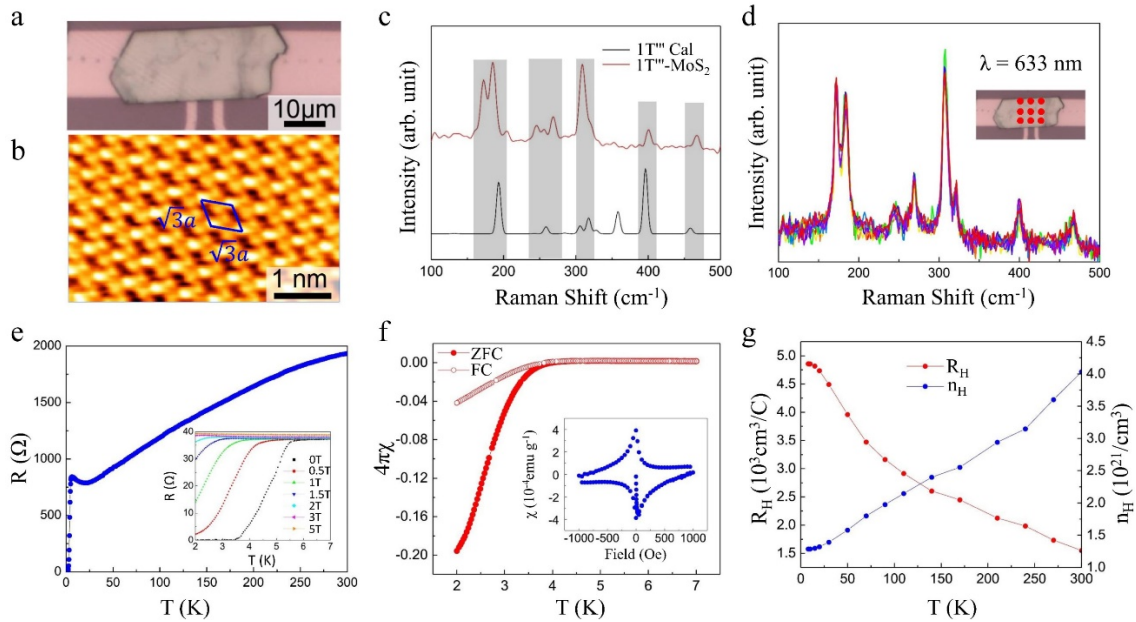

**Figure 2.** Structural characterization and superconductivity of 1T'''-MoS$_2$. (a) The crystal of 1T'''-MoS$_2$ superconductor with electrodes attached for resistance measurements. (b) Atomic-resolution STM image (($V_s$ = 1 V, $I_t$ = 110 pA) taken from 1T'''-MoS$_2$ superconductor. The unit cell with $\sqrt{3}a \times \sqrt{3}a$ superlattice is labelled. (c) Raman spectra of 1T'''-MoS$_2$ superconductor. Calculated Raman spectrum is plotted for comparison, and the grey areas indicate the correspondence of main peaks in theoretical and experimental Raman spectra. (d) High spatial-resolution Raman spectra taken from 1T'''-MoS$_2$. The inset shows the positions where the data were taken. The consistency between them indicates the high quality of 1T'''-MoS$_2$ crystals. (e) Temperature-dependent resistance of 1T'''-MoS$_2$ superconductor. The inset shows the evolution of superconducting transition with applied magnetic field, which indicates the onset T$_c$ =5.3 K at $H$ = 0 Tesla. (f) Magnetic susceptibility of 1T'''-MoS$_2$, from which we estimate that the superconducting volume is about 20%. (g) Hall coefficient and carrier concentration of 1T'''-MoS$_2$ superconducting crystal as a function of temperature, indicating that hole-type carriers are dominant in this material.

Superconductivity in 1T'''-MoS$_2$ is confirmed by transport measurements. In Fig. 2(e), the normal-state resistance exhibits a metallic behavior above T = 25 K, and a moderate upturn shows up below T=25 K with decreasing temperature, and then a superconducting transition occurs at T$_c$ = 5.3 K. In the inset, we show the evolution of low-temperature resistance under various magnetic fields ( $H$ ), and the superconductivity is completely suppressed for $H$ > 2 Tesla. Here, we can clearly see that the resistance drops at T = 5.3 K and reaches to zero at T = 3.5 K for $H$ = 0 Tesla. To evaluate the superconducting volume of the 1T'''-MoS$_2$, we performed the magnetic susceptibility measurement as shown in Fig. 2(f). We estimate that the superconducting volume fraction is about 20% of the total 1T'''-MoS$_2$ product. Moreover, we have carried out Hall measurements to determine the carrier type and the carrier concentration. The Hall data in Fig. 2(g) indicate that the hole-type carriers dominate the transport properties, and the carrier concentration is $4 \times 10^{21}$/cm$^3$ at T = 300 K. Chemical doping and electrostatic doping by ionic gating have been applied to tune the carrier concentration in 2H-MoS$_2$, and the superconducting state exists in a dome-like region in the phase diagram of $T_c$ versus carrier concentration, with the highest $T_c$ = 10.8 K at the optimal doping.[36] In 1T'''-MoS$_2$ superconductor, the carrier concentration is comparable to the doping level of superconducting state induced in 2H-MoS$_2$ as well as high-$T_c$ cuprate superconductors.

Theoretical calculations have predicted that 1T'''-MoS$_2$ is an insulator, and this insulating compound has been synthesized in our studies (the structural and transport properties are shown in Fig. S1 of the Supplemental Material). In order to investigate how the superconductivity emerges in the 1T''' polytype, we have performed various experiments to examine the composition of 1T'''-MoS$_2$. Because sulfuric acid was used in our synthesis procedure, one may question whether there are residual protons in the layered crystals, which can strongly influence the electronic properties. To address this issue, we have performed Thermogravimetry-Mass Spectrum analysis to detect protons and our studies indicate that there are no protons in the 1T'''-MoS$_2$ layered crystals (see the Supplemental Material). Since the precursor material we used for synthesizing 1T''' polytype is K$_x$(H$_2$O)$_y$MoS$_2$, whether there are residual potassium atoms has to be clarified. We have carried out energy-dispersive X-ray spectroscopy (EDS) measurements on a large number of samples. Our EDS measurements show that there is no detectable sign of potassium in 1T''' layered crystals.

In order to examine the difference in electronic states of the superconducting and insulating compounds, we have performed X-ray photoelectron spectroscopy (XPS) and X-ray absorption near edge spectroscopy (XANES) (see Fig. S7 and the discussions in the Supplemental Material).

XANES can provide information on the unoccupied electronic states above the Fermi level, and the stronger spectral weight of the Mo $M_{2,3}$-edge in the superconducting compound is evidence of hole-doping. XPS data show small shifts of both Mo 3d and S 2p core-levels in the superconducting compound compared to the insulating compound, which indicate the chemical potential shift induced by hole doping (see the Supplemental Material for more details). By combining these data and Hall measurements, our studies indicate that the superconductivity can be induced in the 1T'''-MoS$_2$ crystals by hole doping.

By regulating the oxidation procedure (see the Supplemental Material), we have also obtained 1T'-MoS$_2$ layered crystals in the as-grown products. Some trace of superconductivity in 1T'-MoS$_2$ has been observed in polycrystalline materials.[29] However, many uncertainties could arise in the powder samples due to the multiple metastable phases and impurities. Superconductivity in 1T'-MoS$_2$ requires further structural characterizations and transport measurements on high-quality crystals. Fig. 3(a) shows a typical piece of 1T'-MoS$_2$ layered crystal. One may notice that 1T' and 1T''' crystals have different colors, the 1T''' polytype is lighter, while the 1T' polytype appears darker. Indeed, we can empirically distinguish them by color under optical microscope. Fig. 3(b) shows the atomic-resolution STM image taken from 1T'-MoS$_2$, and the characteristic superstructure of a × 2a is marked on the image. Fig. 3(c) shows the temperature-dependent resistance of 1T'-MoS$_2$ layered crystal, which clearly exhibits a superconducting transition at $T_c$ = 4.2 K. This critical temperature is obviously lower than that of 1T'''-MoS$_2$ superconductor. Raman characterization was used to examine the homogeneity of 1T'-MoS$_2$ crystals. As shown in Fig. 3(d), the high spatial-resolution Raman spectra taken from different positions on a piece of crystal are highly consistent with each other, indicating the high quality of 1T'-MoS$_2$ sample, in which we can unambiguously identify the superconductivity of 1T'-MoS$_2$.

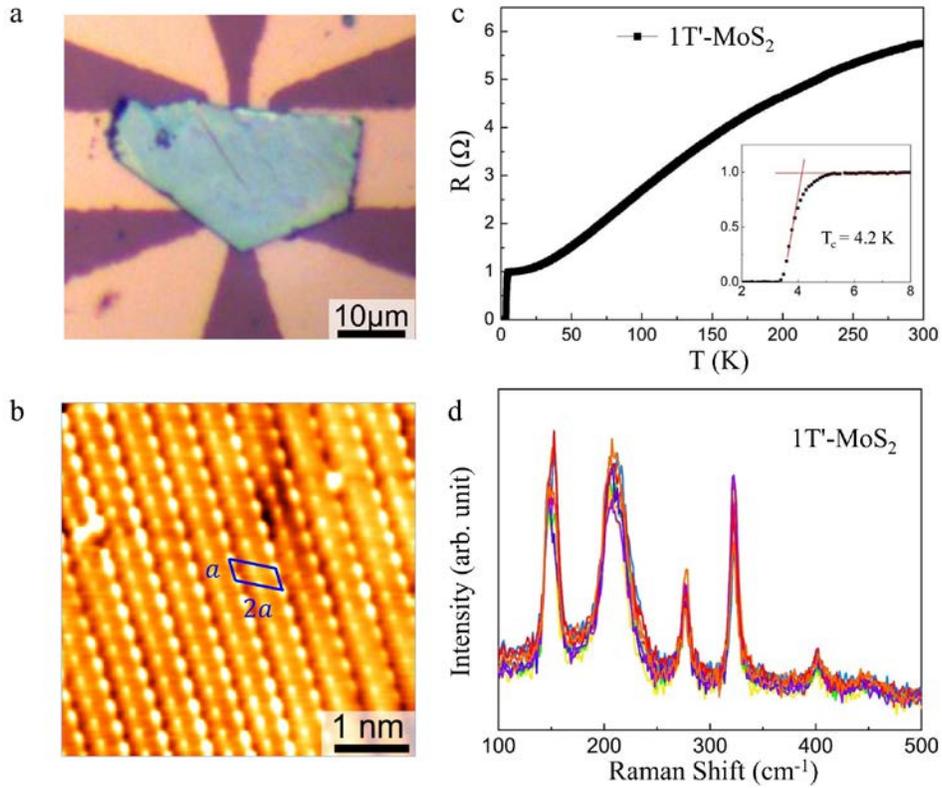

**Figure 3.** Structural characterization and superconductivity of 1T'-MoS$_2$. (a) 1T'-MoS$_2$ crystal with electrodes attached for resistance measurements. (b) Atomic-resolution STM image taken from 1T'-MoS$_2$. The unit cell with a × 2a superlattice is labelled. (c) Temperature dependence of resistance for 1T'-MoS$_2$ crystal. It shows a metallic behavior at high temperature and becomes superconducting with onset transition temperature of 4.2 K as indicated in the inset. (d) High spatial-resolution Raman spectra taken from 1T'-MoS$_2$. The data were taken from different positions of a 1T'-MoS$_2$ crystal, indicating the homogeneity of the layered crystal.

Both of 1T' and 1T'''-MoS$_2$ are metastable. Though their crystal structures can remain for a long time below room temperature, we found that the metastable structures can be transformed into 2H polytype by mild annealing. As shown in Fig. 4(a), when raising the temperature of the 1T'''-MoS$_2$ superconductor to 400 K and keeping the temperature for 10 minutes in He atmosphere, the temperature-dependent resistance of the post-annealed crystal shows a semiconducting behavior, which is consistent with that of the 2H polytype. The corresponding Raman spectra taken before and after annealing demonstrate that the metastable structure of 1T''' polytype is changed to the stable 2H lattice structure (Fig. 4(b)). In Figs. 4(c) and 4(d), we show that the metastable 1T'-MoS$_2$ also exhibits similar structural transition after mild annealing.

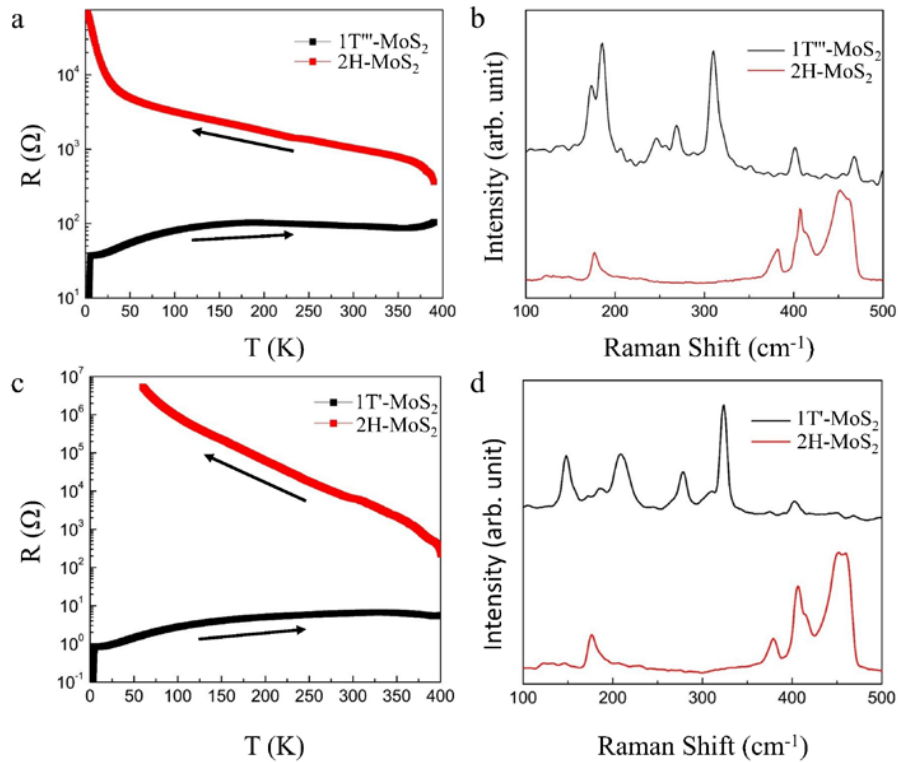

**Figure 4.** The changes from the superconducting and metastable 1T' and 1T'''-MoS$_2$ to the semiconducting 2H-MoS$_2$ by mild annealing. (a) Temperature dependence of resistance for 1T'''

and 2H-MoS$_2$ phases, respectively. The 1T'''-MoS$_2$ crystal shows superconductivity at low temperature. By varying the temperature as indicated by the arrows, the 1T'''-MoS$_2$ crystal can be converted into 2H-MoS$_2$ crystal at T=400 K within 10 minutes in He atmosphere, as evidenced by drastic changes of the temperature-dependent resistance. (b) Raman spectra of 1T''' and 2H-MoS$_2$ taken before and after annealing at T=400 K, confirming the structural transformation. (c) Temperature dependence of resistance for 1T' and 2H-MoS$_2$ phases, respectively. The 1T'-MoS$_2$ crystal shows a superconducting transition at low temperature. The temperature-dependent resistance indicates that a mild annealing at T=400 K in He atmosphere can also convert the 1T'-MoS$_2$ into 2H-MoS$_2$. (d) Raman spectra of 1T' and 2H-MoS$_2$ taken before and after annealing at T=400 K, indicating the structural transformation.

In summary, by using K$_x$(H$_2$O)$_y$MoS$_2$ as the precursor and controlling the oxidation process, we have synthesized high-quality metastable 1T' and 1T'''-MoS$_2$ layered crystals, and their structures and transport properties have been examined, which reveal the intrinsic superconductivity in these materials. Our finding suggests that unusual or unexpected electronic properties, such as superconductivity, exist in these metastable crystals. Versatile atomic configurations of MoS$_2$ and the corresponding variations of physical properties can potentially inspire more extensive investigation of TMDs. For instance, the metastable nature of 1T' and 1T'''-MoS$_2$ crystals can be leveraged to fabricate transistors with other polytypes, and new physics and functionality may take place in the vicinity of the interfaces.

ACKNOWLEDGMENT

This work was supported by the National Key R&D Program of the MOST of China (Grant No. 2017YFA0303001, 2016YFA0300201), and the National Natural Science Foundation of China (Grants No. 11190021, No. 11227902, No. 11534010 and No. 91422303), the "Strategic Priority Research Program (B)" of the Chinese Academy of Sciences (Grant No. XDB04040100), the Hefei Science Center CAS (2016HSC-IU001). We thank Wensheng Yan, Wenhua Zhang, Xusheng Zheng, and Wangsheng Chu for helpful discussions and assistance in characterizations at Photoemission and XMCD endstations in the Hefei National Synchrotron Radiation Laboratory.

* E-mail: zsun@ustc.edu.cn; chenxh@ustc.edu.cn